\documentstyle[12pt]{article}

\begin{document}
\vspace*{0.2in}
\begin{center}
{\Large QCD motivated three-body force and light nuclei} \\
\vspace*{0.3in}
A. A. Usmani
\footnote{pht25aau@amu.ac.in} and F. H. Bhat\\
{\it Department of Physics, Aligarh Muslim University, Aligarh 202 002, India}\\
Afsar Abbas \\
{\it Institute of Physics, Sachivalaya Marg,Bhubaneswar 751 005,
  India} \\ 

\vspace*{0.3cm}
Abstract

\end{center}

\begin{quote}
 The need for the three-body forces (3BF) for reproduction of physical
 properties of light nuclei has been evident for some time. These 3BF are 
doing well for light nuclei, but there seems to be scope for other 3BF 
which may arise from other physical considerations not included so far. Here 
we discuss a QCD motivated new 3BF which should be included in studies of 
nuclei, in particular the light ones. We perform variational Monte Carlo 
calculations including this new 3BF and discuss improvements brought in 
thereby for light nuclei.
\end{quote}

 The study of light nuclei is of special significance, firstly as being 
composed of a few nucleons one hopes that their structure should be simple and
secondly the understanding gained therefrom should be of basic importance for
the study of heavier nuclei. It has therefore been somewhat of a surprise that
these light nuclei themselves present several fundamental problems. It is
however true that at present a rather good understanding has been achieved. 
However, there are still some open issues: (i) Is our
variational wave function, having no explicit dependence 
on quantum chrodynamics (QCD) effects, complete?, (ii) Are
ground-state energies of nuclei sensitive to QCD effects? And if so in what
manner? The answer to these questions, shall be the focus of our attention here.

For since over a decade, it has been known that the best microscopic
calculations using successful two-body forces (2BF) for light nuclei $A=3,4$ 
give underbinding and too large  radii \cite{Gibson}. This hinted at the 
requirement of other forces like the three-body forces (3BF). Once included 
these do improve the fits to the binding energies and the radii of these 
nuclei.

However, one may ask as to all the 3BF included in calculations so 
far\cite{Gibson,Pudliner1997} is it all that one needs? Quite clearly 
this is not correct. There may be other 3BF which have not been included 
so far. Here we shall discuss inclusion of a new 3BF which comes from QCD and
quark model considerations \cite{Abbas86, Abbas88}. Based on basic 
considerations as to confinement in QCD, this new 3BF would be essential in the 
studies of nuclei and in particular the light ones.

We perform a variational Monte Carlo calculation \cite{Pudliner1997, 
Wiringa1991} where we take the Hamiltonian to be
\begin{equation}
H =\sum_{i}^{A}\frac{p_{i}^{2}}{2m}+\sum_{i<j}^{A}v_{ij}+
\sum_{i<j<k}^{A}V_{ijk}.
\end{equation}
Here $v_{ij}$ and $V_{ijk}$ are NN and NNN potentials. For NN potential we use
Argonne $v_{18}$ (AV18) potential \cite{v18} which is written as a sum of 18 
terms, 
\begin{equation}
v_{ij}=\sum_{p=1,18}v_p(r_{ij})O^p_{ij},
\end{equation}
where the first 14 terms are charge independent, 
\begin{eqnarray}
\label{AV14}
O_{ij}^{p=1,14}&=&\left[ 1,\mbox{\boldmath$\sigma$}_i
\cdot\mbox{\boldmath$\sigma$}_j, S_{ij}, {\bf L}\cdot{\bf S}, 
{\bf L}^2, {\bf L}^2\mbox{\boldmath$\sigma$}_i\cdot\mbox{\boldmath$\sigma$}_j,
({\bf L}\cdot{\bf S})^2 \right]\nonumber \\
& &\otimes\left[1,\mbox{\boldmath$\tau$}_i\cdot\mbox{\boldmath$\tau$}_j\right] 
\end{eqnarray}
and the rest,
\begin{equation}
O_{ij}^{p=15,18}=\left[1,  \mbox{\boldmath$\sigma$}_i
\cdot\mbox{\boldmath$\sigma$}_j, S_{ij}\right]\otimes T_{ij},  
(\mbox{\boldmath$\tau$}_{zi} + \mbox{\boldmath$\tau$}_{zj}),
\end{equation}
are charge independence breaking terms. 

For the NNN potential, we have 
\begin{equation}
\label{abp}
V_{ijk}^{TNI}= V_{ijk}^{2\pi}+V_{ijk}^{R}.
\end{equation}
We take Urbana-IX  potential model of the Urbana series 
potentials\cite{Carlson1983, Pudliner1995,Pieper2001}.
The first term is the two- pion exchange attractive component and the second 
is the short but still finite range  phenomenological repulsive component. 
Here, we introduce a new term $V_{ijk}^{Q}$ (more about it a little later) as
\begin{equation}
\label{VQ}
V_{ijk}=V_{ijk}^{TNI}+V_{ijk}^Q,
\end{equation}
which arises from QCD and Quark Model(QM) 
considerations \cite{Abbas86,Abbas88}.  
The variational wave function\cite{Pudliner1997} 
for nuclei may be generalized to write the wave function after properly 
encorporating the $V_{ijk}^Q$ term, 
\begin{eqnarray}
\label{Psi}
\mid\Psi_{\rm v}\rangle &=& 
\left[1+\sum_{i<j<k}^A(U_{ijk}+U_{ijk}^{TNI})
+\sum_{i<j}U_{ij}^{LS}\right]\nonumber \\ 
&&\left[S\prod_{i<j}^A(1+U_{ij})\right]
\mid \Psi_{J}\rangle.
\end{eqnarray}
The $\mid \Psi_J\rangle$ is the antisymmetric Jastrow wave function
\begin{eqnarray}
\label{Psi_J}
\mid \Psi_{J}\rangle& =& 
\left[\prod _{i<j<k}^Af^{Q}_{ijk}\right] 
\left[\prod _{i<j<k}^Af^{c}_{ijk}\right] 
\left[\prod _{i<j}^Af^{c}_{ij}(r_{ij})\right] \nonumber \\
&&{\cal A}\mid \phi_A(JMTT_3)\rangle.
\end{eqnarray}
The functions $f_c(r_{ij})$ and $u_p(r_{ij})$ are the same as given in 
Ref.\cite{Wiringa1991}. The $f^c_{ijk}$ is a three-nucleon correlation 
induced by  $v_{ij}$ which is  discussed in Ref.\cite{Pudliner1997} 
and in references therein. Its explicit mathematical expression is
\begin{equation}
\label{fijkc}
f^c_{ijk}=1+q_1^c 
({\bf r}_{ij}\cdot {\bf r}_{ik})
({\bf r}_{ji}\cdot {\bf r}_{jk})
({\bf r}_{ki}\cdot {\bf r}_{kj})
\mbox{exp}(-q_2^cR_{ijk}).
\end{equation}
Here $R_{ijk}=r_{ij}+r_{jk}+r_{ki}$.  
An optimal variational wavefunction ($\Psi_{\rm v}$) is used to calculate the 
ground-state energy, 
\begin{equation}
E_{\rm v}= \frac{\langle\Psi_{\rm v}\mid H\mid\Psi_{\rm v}\rangle}
{\langle\Psi_{\rm v}\mid\Psi_{\rm v}\rangle} \ge E_0,
\end{equation}
for a  wide spectrum of nuclei, 
nuclear and neutron matter\cite{WiringaRMP93}.

Keeping QCD considerations in mind, the $V_{ijk}^Q$(more about it below)
 gives an infinite repulsion if an accepted  set of spatial configurations 
passes the following conditions operative simultaneously
\begin{equation}
\label{cond1}
r_{ij}\le \lambda_1, r_{jk}\le \lambda_1, r_{ki}\le \lambda_1
\end{equation}
and
\begin{equation}
\label{cond2}
R_{ijk}\le \lambda_2.
\end{equation}
This imposes a corresponding correlation condition on the wave function that 
may be made effective by setting $f_{ijk}^Q$ to zero, thereby setting $\Psi_J$ 
to zero in such case and hence the variational wave function 
$\Psi_{\rm v}$ as given in Eqs. [\ref{Psi},\ref{Psi_J}]. 
The second condition Eq. [\ref{cond2}] is imposed to get a reasonable 
triangular shape of triplets.
Herein, the $\lambda_1$ and $\lambda_2$ are treated as variational parameters 
in the wave function for each nucleus. The $V_{ijk}^Q$ is completely absorbed 
in the wave function. 
Its existence in Eq.[\ref{VQ}], therefore, will be 
decided by the variational principle, which is a sacred law in all variational
calculations.

Now we address as to how $V_{ijk}^{Q}$ arises. As we shall be 
discussing structure of
nuclei in the ground state, we have to ensure that it is the ground state 
structure of nucleons itself that we take into account. Hence we take nucleon 
as consisting of three constituent quarks in the s-state. Note that though the
rms radius of nucleon is taken as 0.8 fm, it is a very diffuse system with 
matter distribution given by $\rho$(r)=$\rho_{0}e^{-\nu r}$.

As the u- and d- quarks are almost degenerate in masses, we take the flavour
symmetry $SU(2)_F$ as a good symmetry, which should be relevant to the study of
ground state properties of nucleus. Hence for our purpose here significant 
group would be  
$SU(12)\supset SU(4)_{SF}\bigotimes SU(3)_{c}$. Here $SU(3)_{c}$ is the QCD 
group and $SU(4)_{SF}\supset SU(2)_{F}\times SU(2)_{S}$, where S denotes spin.
Upto 12 quarks can sit in the s-state for the group SU(12). Colour confinement
is  a property which should follow from the symmetry and dynamics of QCD. 
Therefore such multiquarks built up of 3-, 6-, 9- and 12- quarks should be  
colour singlet objects. The 3- q is our familiar baryon. 
What role do the 6-, 9- and 12- quarks
play in nuclei? In as much as we can successfully study nuclear physics with
nucleonic degrees of freedom the higher multi-quark configuration should play
no role. But it has been found that in physical 
situations where to understand the reality, if the relative distances 
between nucleons  have to be less than 1 fm, then one can not escape 
considering 6- q, 9- q and 12- q configurations \cite{Abbas86,Abbas88,Abbas01}.

For multiquark system (with quark number greater than 3) the concept of hidden
colour plays a basic role. Hidden colour represents that part of the 
multiquark wavefunction which as per the confinement idea of QCD can not be
separated out in terms of physical hadrons and so manifests itself only
inside multiquark systems. If relevant to nuclear physics, this could
represent unique QCD based quark aspects of nuclear physics. 
There have been some claims that hidden colour may not be a useful or unique 
concept for nuclear physics as these may be rearranged in terms of asymptotic 
colour singlet states \cite{Vento}. But as discussed in Ref. \cite{Vento} 
the hidden colour concept is not unique only when the two clusters do not 
overlap strongly and can be separated out asymptotically. However it has been 
demonstrated convincingly that when the 
clusters of 3- q overlap strongly so that the relative distance between 
them goes to zero, the hidden colour concept becomes relevant and unique 
\cite{Abbas86,Abbas88,Vento}. It is only in these situations that we shall 
make use of the concept of hidden colour in our calculation. Ref.\cite{Vento} 
also discusses how antisymmetrization and short range repulsion are related
as per the hidden colour idea at short range. What is important [\cite{Vento}]
is that how near $r=0$ range hidden colour representation is unique.

Group theoretically hidden colour \cite{Sorba} component of 6- q system was 
found to be $80\%$. Next it was found that the 9- q and 12- q components
of hidden colour are predominant i.e. 97.6$\%$ and 99.8$\%$, respectively
\cite{Abbas86,Abbas88}. The $A=3,4$  nuclei $^3$H, $^3$He and $^4$He 
have experimental point-proton radii of 1.60 fm, 1.77 fm and 1.47 fm 
respectively. These radii are obtained by subtracting a proton mean-square 
radius of 0.743 fm$^2$ and $N/Z$ times a neutron mean square radius of 
-0.116 fm$^2$ from the squares of the measured charge radii. 
Given the fact that each nucleon is
itself a rather diffuse object, quite clearly in a size $\leq 1$ fm at the
centre of these nuclei, the three and four nucleons would overlap strongly.
There would be an effective repulsion at the center keeping the three and four 
nucleons away from the centre, as the corresponding 9- and 12- q have 
predominantly hidden colour components. The nucleonic system resists going into 
this configuration. 
This conceptual framework has been found to be useful in making further 
predictions for normal and neutron rich nuclei \cite{Abbas01}.

Let us here concentrate upon the 3BF part. Quite clearly as discussed above,
arguments based on QCD and quark model, indicate that there should be a unique
short range repulsive force in $A=3,4$ nuclei. This is the new $V_{ijk}^Q$ that
we have proposed above.
The $V_{ijk}^Q$ is clearly three body analogue of the 
short range hard core repulsion
in the 2BF \cite{Abbas86,Abbas88,Abbas01}. Just as for 2BF one has hard core 
of infinite repulsion and size $\sim$0.5 fm so here too we implement the 
$V_{ijk}^Q$ as
infinitely repulsive hard core. This hard core is implemented in terms of two 
parameters $\lambda_1$ and $\lambda_2$ (as above)
in a variational Monte Carlo calculations to fit to the physical
quantities of relevance for $A=3,4$ nuclei $^3$H and $^3$He and $^4$He.

\begin{table}
\caption{\label{tab1} 
Variational parameters $\lambda_1$ and $\lambda_2$ in units of fm.}
\begin{center}
\begin{tabular}{l|ccc}
\hline\hline
  & $^4$He($0^+$) &  $^3$He($\frac{1}{2}^+$) & $^3$H($\frac{1}{2}^+$)  \\  
\hline
$\lambda_1$ & 0.55  & 0.63 & 0.62  \\
$\lambda_2$ & 1.2  & 1.5 & 1.5   \\
\hline\hline
\end{tabular}
\end{center}
\end{table}

\begin{table}
\caption{\label{tab2} rms point proton and neutron radii in units of fm.}
\begin{center}
\begin{tabular}{lccc}
\hline\hline
$^AZ(J^\pi)$   & $p$ &  $n$ & $p(Expt.)$  \\  
\hline
$^4$He($0^+)$   & 1.466 & 1.466 & 1.47 \\
$^3$He($\frac{1}{2}^+$) & 1.766 & 1.592 &  1.77\\
$^3$H($\frac{1}{2}^+$)   & 1.613& 1.713 &  1.60\\
\hline\hline
\end{tabular}
\end{center}
\end{table}

\begin{table}
\caption{\label{tab3} 
Energy breakdown for $^4$He, $^3$He and $^3$H with $V_{ijk}^Q$. 
All quantities are in units of MeV.}
\begin{center}
\begin{tabular}{lcccccc}
\hline\hline
$^AZ(J^\pi)$   & $T$ &  $v_{ij}$& $T+v_{ij}$  &  $V_{ijk}^{TNI}$ & $E$ 
& $E({\rm Expt.})$ \\ 
\hline
$^4$He($0^+)$   & 107.44(20) & -129.93(19)& -22.48(2) & -5.28(2) & -27.77(1)  &  -28.30  \\
$^3$He($\frac{1}{2}^+$)  & 48.95(16) & -55.45(16)& -6.50(1) & -1.04(1) & -7.54(1)  &  -7.72  \\
$^3$H($\frac{1}{2}^+$)   & 49.39(23) & -56.69(22)& -7.30(2) & -1.02(2) & -8.32(1)  &  -8.48  \\
\hline\hline
\end{tabular}
\end{center}
\end{table}

In Table \ref{tab1}, we present best values of variational parameters
$\lambda_1$ and $\lambda_2$ tuned to obtain ground-state energies
for the nuclei under consideration.  
For the mirror nuclei: $^3$He and $^3$H whose rms radii are very close to 
each other, $\lambda_1$ is found to be 0.63 fm and 0.62 fm respectively, and  
$\lambda_2$ is found to be about 1.5 fm. 
For the compact nucleus, $^4$He, the values of $\lambda_1$ and $\lambda_2$ 
are smaller which are  found to be 0.55 fm and 1.2 fm. They are down almost 
in the same proportion. 
Thus, $\lambda_1$ and $\lambda_2$ seem to be correlated with the compactness 
of the nucleus.  Our calculations reproduce 
the experimental rms radii which are shown in Table \ref{tab2}. 
In Table \ref{tab3},  the variational energy breakdown
for kinetic energy ($T$), two-body potential energy ($v_{ij}$), total two-body 
energy ($T+v_{ij}$), three-body potential energy ($V_{ijk}$) and  
total energy ($E=T+v_{ij}+V_{ijk}$) with $V_{ijk}^Q$ along with  experimental 
ground state energy are presented. We also report the same without 
$V_{ijk}^Q$ in Table \ref{tab4}.  We note an improvement in the 
upperbound ground-state energy of the nuclei considered herein, by about 
-0.04 MeV, -0.03 MeV and -0.01 MeV in case of $^4$He,$^3$He and $^3$H, 
which may be small but can not be ignored in fine calculations 
while obtaining nuclear spectra through nuclear forces\cite{WP2002}.   
Though the Urbana-IX 3BF is already having a strong phenomenological
short range repulsion, the idea of infinite repulsion still improves
variational results. One may also implement this idea at the level of 
two-body force (2BF) as represented by AV18 in this calculation. Obviously, 
the effect of the infinite repulsion will be more evident with increasing 
number of nucleons coming together, Therefore, 3BF is a better candidate
where QCD and QM considerations should be implemented first.  We aim  to 
implement the same at the level of 2BF later. 
The individual energy pieces $T$, $v_{ij}$ and $V_{ijk}$ are affected 
significantly due to  the presence of $V_{ijk}^Q$.  
We infer that QCD effects are  manifested  through 
$V_{ijk}^Q$ in all the light nuclei considered herein.
We recommend that this force be included in all energy calculations 
for both finite and infinite
nuclear systems, specially for dense and compact systems.

It may be noted that the parameter ($\lambda_1$) above is directly related to
confinement size for a 9-quark system. One knows that so far it has not 
been possible  to obtain the confinement size of proton/neutron from any
fundamental considerations of QCD, hence it would be hopeless to expect to do 
the same for $\lambda_1$. Also as of now since there exists no  definitive
experimental candidates for the tri-baryonic systems, hence it would not be 
possible to lay hands on the ($\lambda_1$) parameter from any phenomenological
quark model perspective. Hence, our method above of treating it as a 
variational fitting parameter.

\begin{table}
\caption{\label{tab4} 
Energy breakdown for $^4$He, $^3$He and $^3$H without $V_{ijk}^Q$. 
All quantities are in units of MeV.}
\begin{center}
\begin{tabular}{lccccc}
\hline\hline
$^A Z(J^\pi)$  & $T$ &  $v_{ij}$& $T+v_{ij}$  &  $V_{ijk}^{TNI}$ & $E$  \\ 
\hline
$^4$He($0^+)$    & 107.94(14)  & -130.45(14) & -22.51(1)  & -5.21(1)  & -27.73(1)   \\
$^3$He($\frac{1}{2}^+$)  & 48.75(15)  & -55.25(16)& -6.50(1)  & -1.01(1)  & -7.51(1) \\
$^3$H($\frac{1}{2}^+$)   & 49.23(23)  & -56.53(23)& -7.28(2)  & -1.03(1)  & -8.31(1)  \\
\hline\hline
\end{tabular}
\end{center}
\end{table}

In Fig. \ref{fig1}, we plot density profiles for $^4$He, $^3$He and $^3$H
in three different panels. The filled circles and  open circles (which appeared
to be strongly overlapping)  represent 
density profiles with  and without $V_{ijk}^Q$, respectively. The cross points
represent the calculations after setting $f_{ijk}^c$=1 and $U_{ijk}^R=0$, 
thereby deleting the central correlation induced by $v_{ij}$ and  short 
range-repulsive correlation from the variational wave function. 
The depression in densities at small r is mainly due to $f_{ijk}^c$ and 
$U_{ijk}^R$ which is a correlation due to $V_{ijk}^R$ in
the wave function.

In summary, we have found that the  QCD motivated new
3BF $V_{ijk}^Q$ proposed herein is essential for the variational wave function
and for the variational ground-state energies.  
Also in $^4$He as per our model there should be an additional short range 
repulsive four-body force which we wish to include in future. 
Results here have been obtained within the framework of successful
variational Monte Carlo calculations.\\

AAU acknowledges  financial support through Grant No. SP/S2/K-32/99 
sanctioned to him under SERC scheme, Department of Science and Technology, 
Government of India. He is also thankful to Robert B. Wiringa for his routines
of s-shell nuclei. FHB acknowledges a junior research fellowship from Council
of Scientific and Industrial Research, Government of India.


\begin{thebibliography}{299}
\bibitem{Gibson}B. F. Gibson, Nucl. Phys. A {\bf543}, 1c(1992). 
\bibitem{Pudliner1997}B. S. Pudliner, V. R. Pandharipande, J. Carlson, 
                 S. C. Pieper and R.B. Wiringa,
                 Phys. Rev. C {\bf56}, 1720(1997).
\bibitem{Abbas86}A. Abbas, Phys. Lett. B {\bf167} (1986) 150
\bibitem{Abbas88}A. Abbas, Prog. Part. Nucl. Phys. {\bf20} (1988) 181
\bibitem{Wiringa1991}R. B. Wiringa, Phys. Rev.  C {\bf 43}, 1585(1991).
\bibitem{v18}R. B. Wiringa, V. J. G. Stoks and R. Schiavilla, Phys. Rev.
         C {\bf 51}, 38(1995).
\bibitem{Carlson1983}J. Carlson, V. R. Pandharipande and R. B. Wiringa, 
          {\bf A401}, 59(1983).
\bibitem{Pudliner1995}B. S. Pudliner, V. R. Pandharipande, J. Carlson, 
                 and R.B. Wiringa, Phys. Rev. Lett. {\bf74}, 4396(1995).
\bibitem{Pieper2001}S. C. Pieper, V. R. Pandharipande, 
         R.B. Wiringa and J. Carlson, Phys. Rev. C {\bf64} 014001(2001).
\bibitem{WiringaRMP93} R. B. Wiringa, Rev. Mod. Phys. {\bf 65}, 231(1993).
\bibitem{Abbas01}A. Abbas, Mod. Phys. Lett. A {\bf16}, 255(2001).
\bibitem{Vento}P. Gonzalez and V. Vento, Il Nuovo Cimento {\bf A106}, 795(1992).
\bibitem{Sorba}V. A. Matveev and P. Sorba,  Lett. Nuovo Cimento {\bf20}, 
          435(1977). 
\bibitem{WP2002} R.B. Wiringa and S. C. Pieper, Phys. Rev. Lett. {\bf89}, 
                182501(2002).


\end{thebibliography}
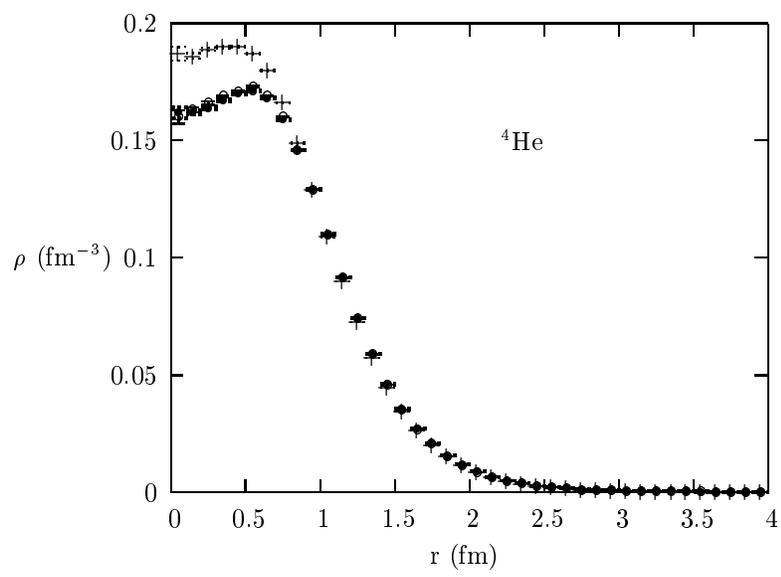
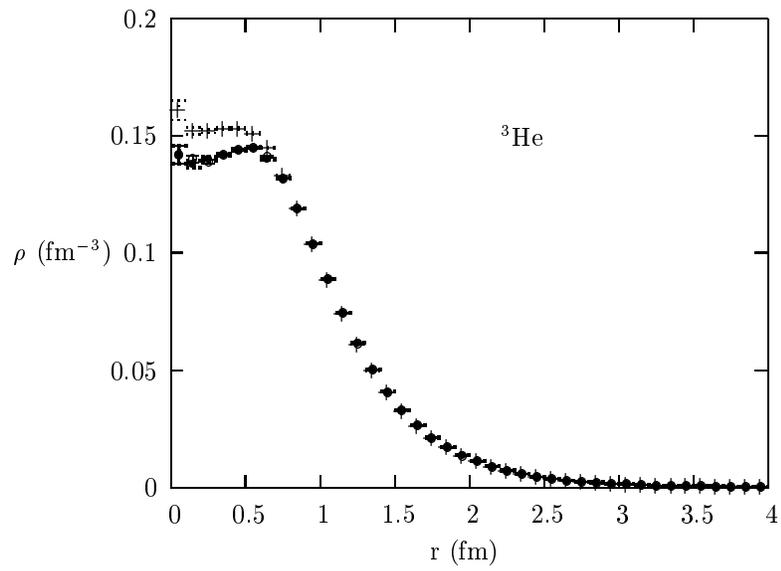
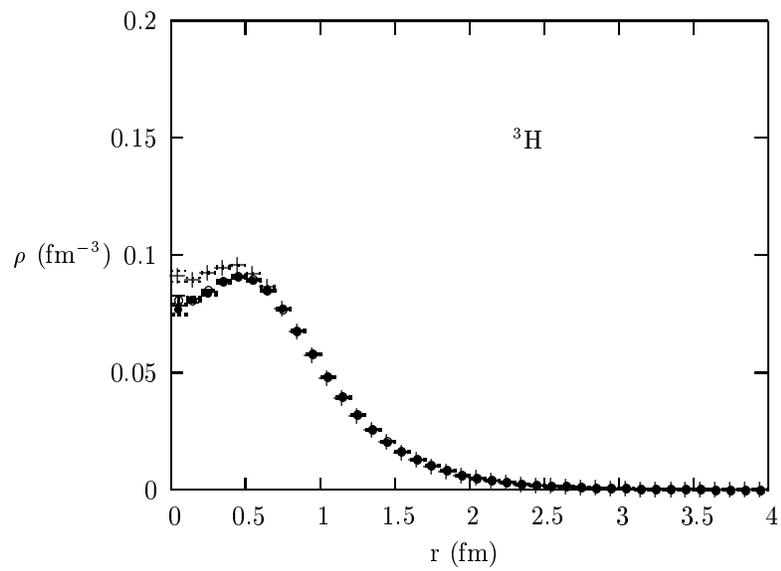
\end{document}